\documentstyle[aps]{revtex}
\begin{document}
\draft

\title{Prospects for p-wave paired BCS states of fermionic atoms}
\author{L. You and M. Marinescu}
\address{School of Physics, Georgia Institute of Technology,
Atlanta, GA 30332-0430}
\date{\today}
\maketitle

\begin{abstract}
We present theoretical prospects for creating p-wave
paired BCS states of magnetic trapped fermionic atoms.
Based on our earlier proposal of using dc electric fields
to control both the strength and anisotropic characteristic
of atom-atom interaction and our recently completed multi-channel
atomic collision calculations we discover that p-wave pairing
with $^{40}$K and $^{82,84,86}$Rb in the low field seeking maximum
spin polarized state represent excellent choices for achieving
superfluid BCS states; and may be realizable with current
technology in laser cooling, magnetic trapping, and evaporative/sympathetic
cooling, provided the required strong electric field can be applied.
We also comment on the prospects of similar p-wave paired BCS states
in $^{6}$Li, and more generally on creating other types exotic BCS states.
Our study will open a new area in the vigorous pursuit to
create a quantum degenerate fermionic atom vapor.

\end{abstract}


\narrowtext
The success of atomic Bose-Einstein condensation (BEC) \cite{bec}
has induced an exponential growth of interest in the properties
of ultracold dilute quantum gases. Of particular interest now is the
physics of trapping and cooling of fermionic atoms \cite{italy}.
Indeed the prospect of superfluidity with dilute atomic
vapors has already been studied by several
groups \cite{Stoof,modawi,Kagan,baranov}.

In this Letter we present a theoretical study of the prospect
of superfluidity in magnetic trapped fermionic atoms. We conduct our
discussions for the three alkali species currently being
investigated at several labs: $^6$Li ($I=1$), $^{40}$K ($I=4$),
and $^{82,84,86}$Rb ($I=1,2,2$) \cite{lanl}. This Letter is organized as
follows: We briefly review the existing proposals
for BCS states of magnetically trapped atoms \cite{Stoof,modawi,Kagan},
after which we present our proposal based on the use of external
dc-electric (dc-E) fields to induce anisotropic atom-atom
interactions \cite{mm}. We then address its advantages and
present results of our detailed multi-channel atomic collision
calculations. We conclude with a discussion of other possibilities
of BCS states offered by our proposal, notably,
the analogies of the Anderson-Brinkman-Morel \cite{am} and
Balian-Werthamer \cite{bw} states in $^3$He \cite{Lee,tony}.

In the standard BCS theory \cite{bcs} of superconductivity
the attractive interaction between electrons of opposite spin and
momentum causes the (Cooper) instability of the filled Fermi sea
ground state. Physically, this effect
can be roughly understood as some kind of condensation of ``bosonic''
Cooper pairs \cite{tony,bcs}. In the weak coupling limit when
$k_BT_c \ll \hbar\omega_D$, ($\hbar\omega_D$ is a characteristic
energy over which an attractive interaction persists around the Fermi
surface), the transition temperature $T_c$ is
\begin{eqnarray}
k_BT_c\sim \hbar\omega_D \exp \left( -{1\over N(0) |V|} \right).
\label{bcst}
\end{eqnarray}
$N(0)$ is the electronic density of states at the Fermi surface,
and is $\propto k_F$, the Fermi momentum. $V$ is the two body
scattering amplitude, assumed constant and negative around the
Fermi surface. Although derived for the case of a homogeneous system,
as argued in several recent studies \cite{Stoof,baranov}
Eq. (\ref{bcst})
should also provide good estimates for current magnetic traps where
the local density approximation is expected to be valid.
We note that the computation of $V$ is a complicated
and difficult many-body problem, and reliable values can rarely
be obtained. For an ultracold weakly interacting dilute atom gas,
the situation is quite different, the scattering amplitude
can be accurately computed, and is usually dominated by the lowest
several partial waves (at sub-mini-Kelvin temperatures).
The effective interaction $V_l$ for the dominant l-th
partial wave term is $\sim \delta_l(k_F)/k_F$, where $\delta_l(k_F)$
is the phase shift for the l-th partial wave.

In the pairing of two fermions the Pauli exclusion
principle requires the total wavefunction to be antisymmetric.
Therefore the orbital angular momentum $l$ has to be
odd for spin symmetric pairs, and even for spin asymmetric ones.
For spherically symmetric interatomic potentials which
vanish exponentially,
$\delta_l(k)\sim k^{2l+1}$ at low energies when $k\to 0$.
For a potential of the asymptotic form $-1/R^n$ ($n>4$),
the phase shift scales as $\delta_l(k)\sim k^{2l+1}$
for $l<(n-3)/2$ and as $\delta_l(k)\sim k^{n-2}$ for $l>(n-3)/2$.
At a large internuclear distance $R$, the typical ground state
interatomic potential is asymptotically dominated by the
van der Waals term $-C_6/R^6$,
where $C_6$ is the dispersion coefficient.
Therefore the phase shift scales as $\delta_l(k)\sim k^{2l+1}$
for $l=0,1$ and as $\delta_l(k)\sim k^{4}$ for $l\ge 2$.
The only significant term is then from the s-wave,
described by the
scattering length $s_{\rm sc}\equiv -\lim_{k\to 0}\delta_0(k)/k$.

Based on the observation above, one realizes that
atoms in spin symmetric states interact only
through a vanishingly small p-wave ($\sim k^3$).
For $^6$Li in the triplet (electronic spin symmetric) state,
the p-wave scattering length defined as
$a_{\rm sc-p}^3\equiv\lim_{k\to 0} -{\delta_1(k)/k^3}$ \cite{randy},
is about $-35$ (a.u.), corresponds to a very weak attraction.
While the s-wave triplet scattering length $a_{\rm sc}$
is enormous at about $\sim -2160$ (a.u.) due to
a shallow bound state \cite{randy}.
In a recent paper Stoof {\it et. al.} studied the prospects for
superfluidity of magnetically trapped $^6$Li \cite{Stoof}. In the
maximum polarized state $|6\rangle$,
a p-wave BCS pairing occurs at a transition temperature of
\begin{eqnarray}
T_c\sim {\epsilon_F\over k_B}
\exp \left[-{\pi\over {2(k_F |a_{\rm sc-p}|)^3}}\right],
\end{eqnarray}
too slow to be practical. For instance,
with a density of $10^{12}$/cm$^3$, we have
$\epsilon_F/k_B\sim 600$ (nK) and
$k_F |a_{\rm sc-p}|\sim 7\times 10^{-3}$ \cite{Stoof}.

Kagan et. al. suggested the trapping of atoms in multiple
hyperfine states \cite{Kagan}. They discovered that two atoms
in the same state ($|6\rangle$ of $^6$Li)
can attract through phonon induced interactions
(caused by density fluctuations in the other state,
$|5\rangle$ of $^6$Li), and pairing occurs at
\begin{eqnarray}
T_c\sim {\epsilon_F\over k_B}
\exp \left[-13\left({\pi\over {2k_F |a_{\rm sc}|}}\right)^2\right],
\end{eqnarray}
where the large s-wave scattering length (for $^6$Li) enters.
Despite a dramatic enhancement (from the $k_F$ dependence),
this scheme is still difficult to realize.
For a density of about $10^{12}$/cm$^3$
per state, (a total density of $2\times 10^{12}$/cm$^3$),
one gets $k_F|a_{\rm sc}|\sim 0.4$. The prefactor $13$ in
the exponent dictates an excessively low temperature.

Recently Stoof {\it et. al.} \cite{Stoof} proposed an
s-wave pairing of atoms in different hyperfine states
($|5\rangle$ and $|6\rangle$ of $^6$Li) which occurs at
\begin{eqnarray}
T_c\sim {\epsilon_F\over k_B}
\exp \left(-{\pi\over {2k_F |a_{\rm sc}|}}\right).
\label{tm}
\end{eqnarray}
For a density of $\sim 10^{12}$/cm$^3$, $T_c$
is about a few tens of a nano Kelvin,
potentially within experimental reach if the atomic decay
channels are suppressed \cite{jin}. There are, however, major obstacles
towards trapping more than one hyperfine state simultaneously:
first, the two body exchange loss channel is now open
and the dipolar rates generally also increase; second,
for stable BCS states, the pairing energies for atoms in
different hyperfine states have to overcome the Zeeman energy deficit.
An interesting proposal was also recently suggested by
Modawi et. al. \cite{modawi} utilizing the s-wave pairing of
atoms in spin anti-symmetric states with a result
similar to Eq. (\ref{tm}).

In a recent letter \cite{mm} we proposed to control
atom-atom interactions at ultra-low temperatures with a dc-E.
In the presence of a dc-E the spherical
symmetry of the interacting atoms is distorted, consequently
the long-range inter-atomic potential is modified by the
addition of an anisotropic dipole interaction term
\begin{equation}
V_E(R)=-\frac{C_E}{R^3} P_2(\cos\theta),
\label{ve}
\end{equation}
where $C_E=2{\cal E}^2\alpha_1^A(0)\alpha_1^{B}(0)$,
$\cal E$ is the strength of the dc-E and $\alpha_1^{A,B}(0)$ are
the static atomic dipole polarizabilities for the two atoms
denoted by $A$ and $B$. $P_2$ is the
Legendre polynomial of order 2 and $\theta$ is the angle between
the directions of the electric field and the internuclear axis.
A complete low energy scattering treatment reveals that the total two
body (unsymmetrized) scattering amplitude takes the form \cite{mm}
\begin{equation}
f({\vec k},{\hat k'})=4\pi\sum_{lm,l'm'}
t_{lm}^{l'm'}(k)Y_{lm}^*({\hat k})Y_{l'm'}({\hat k'}).
\label{a6}
\end{equation}
The reduced T-matrix elements
$t_{lm}^{l'm'}\equiv \lim_{k\to 0}T_{lm}^{l'm'}/k$, (which has
a physical meaning similar to that of the s-wave scattering
length $a_{\rm sc}$),
are all nonvanishing asymptotically a $k\to 0$.
One can intuitively understand this property
($T_{lm}^{l'm'} \sim k$ in the low energy limit) based on
the discussion of the $k$-dependence of the phase shifts
of an asymptotic potential $-1/R^n$.
We found that the couplings between different channels
are essentially given by $V_2\sim1/R^3$.
Thus, for each channel, the effective potential generated by the couplings,
behaves in the limit of large $R$ as $1/R^6$. Therefore the character
of the asymptotic $R$-dependence
of the total effective potential for each partial wave equation
is decided by the diagonal part of the total potential. For $l=0$
the effective potential behaves as $1/R^6$ and so $\delta_0\sim k$.
For $l\neq0$ the effective potentials behave as $1/R^3$
and so $\delta_l\sim k$. The anisotropic nature required
new numerical techniques to ensure the stability
of the low energy scattering calculations. Detailed
discussions of the complete T-matrix elements for all alkali
metal atoms will be reported elsewhere \cite{newmm}.

Both the magnitude and sign of the $t_{lm}^{l'm'}$ are tunable
by changing ${\cal E}$ \cite{mm}.
Quite generally, we found that within the same $l$ and $l'$
manifold the $m=m'=0$ is always the largest $t_{lm}^{l'm'}$ term,
and the larger $m$ and $m'$, the smaller $t_{lm}^{l'm'}$.
Away from resonances, smaller $l$ and $l'$ correspond to
larger $t_{lm}^{l'm'}$.
Such a situation allows for the general
pairing schemes as in $^3$He \cite{am,bw,tony}.
One obtain the $T=0$ gap equation (in the weak coupling limit),
\begin{eqnarray}
{\cal C}(\vec k)={4\pi\hbar^2\over M}\sum_{\vec k'}\sum_{lm}\sum_{l'm'}
(4\pi)\,t_{lm}^{l'm'}Y_{lm}^*(\hat k)Y_{l'm'}(\hat k')
{ {\cal C}(\vec k') \over 2E_{\vec k'}},
\label{cg}
\end{eqnarray}
where the general quasi-particle excitation spectra is
\begin{eqnarray}
E_{\vec k}=\sqrt{\epsilon_{\vec k}^2+|{\cal C}(\vec k)|^2}\ ,
\end{eqnarray}
for the homogeneous case.
$\epsilon_{\vec k}={\hbar^2k^2\over 2M}-\epsilon_F$
is the bare particle energy (measured from the Fermi surface
at $\epsilon_F$). Such a general pairing scheme results
in an anisotropic gap ${\cal C}(\vec k)$ and the excitation
spectra $E_{\vec k}$ \cite{am,bw,tony}.
In particular when one of the $t_{lm}^{lm}$ is made
dominant and attractive (positive), we can simplify the analysis
by assuming
\begin{eqnarray}
{\cal C}_{lm}(\vec k)\approx\Delta_{lm} Y_{lm}^*(\hat k).
\end{eqnarray}
One then obtains the ($T\ne 0$) gap equation
\begin{eqnarray}
1={4\pi \hbar^2\over M}\sum_{\vec k'}(4\pi)\,t_{lm}^{lm}|Y_{lm}(\hat k')|^2
{1\over 2E_{\vec k}} \tanh \left({1\over 2}
\beta E_{\vec k} \right),
\label{gt}
\end{eqnarray}
with $\beta=1/k_B T$. At $T_c$, $\Delta_{lm}(T_c)=0$,
Eq. (\ref{gt}) can be approximately solved to give a
result similar to (\ref{bcst}),
\begin{eqnarray}
T_c\sim {\epsilon_F\over k_B}
\exp \left(-{\pi\over {2k_F |t_{10}^{10}|}}\right).
\label{tc}
\end{eqnarray}
except for the substitution
of $V\to -t_{lm}^{lm}$.
The case of a p-wave pairing with $t_{10}^{10}$ being the
dominant (and positive) term is indeed true
for $^{6}$Li, $^{40}$K, and $^{82,84,86}$Rb in the
maximum polarized state.
These paired superfluid states are the analogues of
the $^3$He in the $A_1$ phase.

Such a scheme is interesting since pairing involves
only atoms in the same hyperfine state,
yet the strength of the pairing can be as strong as that
for a s-wave pairing in different hyperfine states
\cite{Stoof,modawi,ho}. This greatly increases
the critical temperature for the BCS pairing to occur.
The actual values of $t_{lm}^{l'm'}$
are illustrated by our calculations
in Figs. \ref{fig1}, \ref{fig2}, and \ref{fig3}
for $^6$Li, $^{40}$K, and $^{82,84,86}$Rb respectively.
Although hyperfine structures are not included in these
calculations presented. We expect they give the
correct order of magnitude and the ${\cal E}$-dependence
since collisions among atoms in the maximum polarized
state mainly proceed along the triplet potential curves.
We have also estimated the critical temperatures Eq. (\ref{tc})
for typical trap parameters. They are plotted in dashed lines
with respect to the logarithmic vertical axis to the right.
The number density at the trap center is higher than the
$10^{12}$/cm$^3$ used for previous estimates, justified since
all inelastic processes would be slower in our scheme, the
same reason why the BEC was originally realized in
similar maximum polarized states. In the maximum polarized
states, the spin exchange collisional loss is eliminated
and the two-body dipolar rates are suppressed. The three body loss
inside an electric field is a much more complicated issue
currently under investigation. This rate seems highly reasonable
since the centrifugal barrier in the p-wave collision channel prevents
close encounters of two atoms at short distances.

More generally, one may consider possibilities of
pairing with $t_{1\pm1}^{1\pm1}$ terms if resonance structures
can be induced. Such resonances may also provide larger
$t_{lm}^{l'm'}$ values at smaller dc-E fields.
We are currently pursuing detailed atomic
collision calculations including hyperfine structures inside a dc-E
to answer these questions.
Inside an optical trap \cite{mito},
different Zeeman states of the same hyperfine spin state are
degenerate, but the effect of the dc-E can still be induced.
In such a case one would expect that similar to the analysis of
the Anderson-Brinkman-Morel states and
Balian-Werthamer states in $^3$He \cite{am,bw,tony}, one can
consider all kinds of spin symmetries of the paired atoms \cite{ho}.
This would represent a more interesting scenario than in the
case of $^3$He because the total spins can be much larger, e.g.
for $^{40}$K, $f=9/2$ and $7/2$ respectively in the
two ground hyperfine manifolds \cite{ho}.

As illustrated in Fig. \ref{fig1} for $^6$Li, within the range of
dc-E field considered, the induced p-wave interaction is too weak
for the BCS pairing to occur at reasonable temperatures. This
is a result of the small electric polarizability of $^6$Li.
However, for $^{40}$K (Fig. \ref{fig2}) and $^{82,84,86}$Rb
(Fig. \ref{fig3}), pairing can be induced to occur at
temperatures $\sim\hbar\omega_r/k_B$ for dc-E field of
about 3,000 (kV/cm) and 1,000 (kV/cm) respectively. This should
represent a reasonable prospect to aim for within current
experimental capabilities.
The estimate of $T_c$ is based
on the numbers for currently available cylindrical magnetic traps
(with $\omega_z=\lambda\omega_r$, where $\omega_z$ and $\omega_r$
are respectively the axial and radial frequencies), i.e. with the width of
the trap ground state $a_r=\sqrt{\hbar/2M\omega_r}\gg |t_{lm}^{lm}|$
\cite{butts,jin}.
The Fermi energy and the momentum are respectively
$\epsilon_F=\hbar\omega_r (6\lambda N)^{1/3}$ and
$k_F=(6\lambda N)^{1/6}/a_r$ for a noninteracting fermi gas in these traps.
The effective spatial density at the trap center is
$n(\vec r=0)\approx (\lambda N/6)^{1/3}/(\pi^2 a_r^3)$.
Although the required electric field is high for a dc-E configuration,
such field strengths can be easily achieved with lasers
(e.g, CO$_2$ laser) in a quasi-static situation \cite{noteco2}.

In conclusion, we have proposed a BCS pairing scheme for
atoms inside an external dc-E. We have performed detailed atomic
calculations demonstrating the strong anisotropic pairing
interactions for polarized atoms.
By turning the p-wave scattering amplitude to larger values,
we can effectively tune (higher) the critical temperature $T_c$.
Compared with the existing proposals, our scheme
seems to have several distinct advantages:
First, the pairing only involves atoms in the maximum polarized
state. This allows the same phase space density to be reached at a
lower spatial density, thus lowering three-body decays.
Second, trapping inside the maximum polarized triplet
hyperfine states completely eliminates the large (usually
dominant) two-body decay mechanism: those due to
spin exchange collisions. Additionally, it suppresses
the two-body dipolar decay processes since the final decay
channels are more restricted. Third, the applied dc-E provides
a tuning knob for optimal experimental control.
All indications seem to
be consistent with our proposal at the moment. We therefore encourage
experimentalists to attempt the implementation of
the strong electric fields required. It may also become possible
to utilize the low energy scattering resonances to induce
dominant two-body attractions in d-, e-, and maybe even f-waves.
If the fermionic gases can be confined in an optical dipole trap
similar to those recently realized for BEC at MIT \cite{mito},
one may be able to create more exotic BCS states such as those
of the A- and B- phases in $^3$He, but with richer
spin structures \cite{modawi,ho}.

This work is supported by the ONR grant 14-97-1-0633
and by the NSF grant No. PHY-9722410. We want to thank Drs.
R. Hulet and C. Green for providing triplet potentials
for Li and Rb respectively.
Most of the work was completed while L.Y. was a long
term visitor to the Institute for Theoretical Physics (ITP).
L.Y. would like to thank many enlightening discussions
with participants of the ITP BEC workshop.
In particular, he would like to thank Profs. J. Ho, M. Gunn, Y. Kagan,
and A. J. Leggett for the discussion of BCS physics. He
would also like to thank the ITP for its hospitality and the support
of the NSF grant No. PHY94-07194 (to ITP).

\appendix
\section{Scattering from a anisotropic potential}
In this appendix we provide the essential formulation we
developed for the scattering from the anisotropic
potential containing the electric induced
dipole-dipole interaction. As was shown in \cite{mm},
the atom-atom interaction potential inside a dc
electric field of strength ${\cal E}$ is
\begin{eqnarray}
V({\vec R})=V_0(R)+V_E(R),
\label{av}
\end{eqnarray}
where $V_0$ is the symmetric part (for the usual
collisional studies), and the induced anisotropic part is
\begin{equation}
V_E(R)=-\frac{C_E}{R^3} P_2(\cos\theta),
\end{equation}
where $C_E=2{\cal E}^2\alpha_1^A(0)\alpha_1^{B}(0)$ is the
electric induced dipole interaction coefficient and $\alpha_1^{A(B)}(0)$
are the static atomic dipole polarizabilities of atom A and B respectively.
$P_2(.)$ is the Legendre polynomial of order 2 and $\theta$ is the angle
between the directions of the electric field and the internuclear axis.

With such a potential (\ref{av}), the usual partial wave expansion
has to be modified. Consistent with the standard definition for the
scattering amplitude, we denote
the incident momentum (relative motion of the two atoms) by
$\vec k$, then the scattering wave function to be computed is
\begin{eqnarray}
\psi(\vec r)\sim e^{i\vec k\cdot\vec r}+{f(\vec k,\hat r)\over r}e^{ikr},
\end{eqnarray}
note, here $\vec r$ is the relative coordinates between the
two nucleus. The scattering is described by $f(\vec k,\hat r)$
which (for elastic scattering) is always on the energy shell,
i.e. the scattering momentum $\vec k'=k\hat r$. We also expand
the incident wave according to
\begin{eqnarray}
e^{i\vec k\cdot \vec r}=4\pi \sum_{lm} i^l j_l(kr) Y_{lm}^*(\hat k)
Y_{lm}(\hat r),
\end{eqnarray}
with the asymptotic expansion,
\begin{eqnarray}
j_l(kr) \sim {1\over kr}\sin (kr -l{\pi\over 2}), \hskip 36pt r\to\infty.
\end{eqnarray}
Therefore
\begin{eqnarray}
e^{i\vec k\cdot \vec r}\sim
{4\pi\over kr} \sum_{lm} i^l \sin (kr -l{\pi\over 2}) Y_{lm}^*(\hat k)
Y_{lm}(\hat r).
\end{eqnarray}
The scattering amplitude $f(\vec k,\hat r)$ can also
be expanded onto the (complete)
basis $Y_{lm}(\hat r)$ to yield
\begin{eqnarray}
f(\vec k,\hat r)={4\pi\over k}\sum_{lm} T_{lm}(\vec k)Y_{lm}(\hat r).
\end{eqnarray}

We then have for the scattering solution
\begin{eqnarray}
r\psi_{\vec k}(\vec r)=\phi_{\vec k}(\vec r)
={4\pi\over k}\sum_{lm}i^l\left[Y_{lm}^*(\hat k)\sin (kr -l{\pi\over 2})
+ T_{lm}(\vec k)e^{ikr-il{\pi\over 2}}\right]Y_{lm}(\hat r).
\label{bc}
\end{eqnarray}

Therefore the solution $\phi_{\vec k}(\vec r)$ can also be
solved in the multi-channel $Y_{lm}(\hat r)$ basis.
We have the coupled equation,
\begin{eqnarray}
H_l\phi_{lm}(\vec r)=\sum_{l'm'}i^{l'-l}
\langle lm|V(\vec r)|l'm'\rangle \phi_{l'm'},
\end{eqnarray}
where here $H_l$ is the effective Hamiltonian for the l-th partial
wave
\begin{eqnarray}
h_l&&=-{d^2\over dr^2}+{l(l+1)\over r^2}.
\end{eqnarray}

The boundary conditions for the scattering are as from
Eq. (\ref{bc})
\begin{eqnarray}
\phi_{lm}\sim Y_{lm}^*(\hat k)\sin (kr -l{\pi\over 2})
+ T_{lm}(\vec k)e^{ikr-il{\pi\over 2}}.
\end{eqnarray}

Rewrite the total potential as
\begin{eqnarray}
V(\vec r)=V_0+V_2\sqrt{4\pi}Y_{20}(\hat r),
\end{eqnarray}
with
\begin{eqnarray}
V_2(\vec r)=-{C_3\over r^3}{1\over \sqrt{5}},
\end{eqnarray}
we obtain
\begin{eqnarray}
\langle lm|V(\vec r)|l'm'\rangle=V_0\delta_{ll'}\delta_{mm'}
+\delta_{mm'}V_2\langle lm|20|l'm\rangle,
\end{eqnarray}
and the final multi-channel form for the scattering equations
\begin{eqnarray}
h_l\phi_{lm}(r)=V_0\phi_{lm}(r)+\sum_{l'}i^{l'-l}
\langle lm|V(\vec r)|l'm'\rangle \phi_{l'm}(r).
\end{eqnarray}

Due to the symmetriy of the matrix element
$\langle lm|20|l'm'\rangle=0$, if $l+l'+2=$odd.
We immediately see that the even and odd parity (l)
channels decouple. Specifically, if we keep open
only the lowest two channels, we obtain
\begin{eqnarray}
\left\{ \begin{array}{l}
h_0\phi_{00}(r)=V_0 \phi_{00}(r)-V_2\phi_{20}(r),\\
h_2\phi_{20}(r)=(V_0+{2\over 7}\sqrt{5}V_2)\phi_{20}(r)-V_2\phi_{00}(r),
                            \end{array} \right .
\end{eqnarray}
and
\begin{eqnarray}
\left\{ \begin{array}{l}
h_1\phi_{10}(r)=(V_0+{2\over\sqrt{5}}V_2)\phi_{10}(r)
-3\sqrt{3\over 35}V_2\phi_{30}(r),\\
h_3\phi_{30}(r)=(V_0+{4\over 3\sqrt{5}}V_2)\phi_{30}(r)
-3\sqrt{3\over 35}V_2\phi_{10}(r),
                            \end{array} \right .
\end{eqnarray}
\begin{eqnarray}
\left\{ \begin{array}{l}
h_1\phi_{1\pm1}(r)=(V_0-{1\over\sqrt{5}}V_2)\phi_{1\pm1}(r)
-3\sqrt{2\over 35}V_2\phi_{3\pm1}(r),\\
h_3\phi_{3\pm1}(r)=(V_0-{1\over\sqrt{5}}V_2)\phi_{3\pm1}(r)
-3\sqrt{2\over 35}V_2\phi_{1\pm1}(r).
                            \end{array} \right .
\end{eqnarray}

To match the boundary conditions, we further expand the matrix
scattering element according to
\begin{eqnarray}
T_{lm}(\vec k)/k=\sum_{l'm'}t_{lm}^{l'm'}(k)Y_{l'm'}(\hat k).
\end{eqnarray}
As discussed in detail in \cite{mm}, the nature of the
asymptotic dipole-dipole interaction potential is such that
all $t_{lm}^{l'm'}(k)$ become independent of $k$ at low
energies.
Upon matching with the numerically integrated solutions,
we obtain all $t_{lm}^{l'm'}(k)$.

The total scattering amplitde is therefore
\begin{eqnarray}
f(\vec k,\hat r)=
f_{{\rm even}\ ll'}(\vec k,\hat r)+f_{{\rm odd}\ ll'}(\vec k,\hat r).
\end{eqnarray}

For identical particle scattering one has to worry about
the symmetrized (anti-symmetrized) amplitude
$f_{\rm S}$ ($f_{\rm A}$) for bosons (fermions).
In the spin triplet eletronic state of two atoms, one has
\begin{eqnarray}
f_{\rm S}(\vec k,\hat r)&&={1\over \sqrt 2}
[f(\vec k,\hat r)+f(\vec k,-\hat r)]\nonumber\\
&&=\sqrt{2}f_{{\rm even}\ ll'}(\vec k,\hat r),
\end{eqnarray}
and
\begin{eqnarray}
f_{\rm A}(\vec k,\hat r)&&={1\over \sqrt 2}
[f(\vec k,\hat r)-f(\vec k,-\hat r)]\nonumber\\
&&=\sqrt{2}f_{{\rm odd}\ ll'}(\vec k,\hat r).
\end{eqnarray}

\begin{figure}
\caption{For $^{6}$Li with $\lambda=1$, $N=10^6$,
and $\omega_r= (2\pi) 400$ (Hz). We then have $a_r=1.45$ ($\mu$m),
$\epsilon_F\approx 181.7$ ($\hbar\omega_r$), $k_F\approx 13.5/a_r$,
and $n(\vec r=0)=1.36\times 10^{13}$/cm$^3$. The solid lines denote
$t_{lm}^{lm}$ in (a.u.) and refer to the
left vertical scale, while the dashed line denotes $k_BT_c$
(for pairing with $t_{10}^{10}$) in
units of $\hbar\omega_{r}$ and refers to the
(10 based) logarithmic right vertical scale.
}
\label{fig1}
\end{figure}

\begin{figure}
\caption{The same as for Fig. 1, but now for $^{40}$K with
$\lambda=0.1$, $N=10^6$,
and $\omega_r= (2\pi) 400$ (Hz). We then have $a_r=0.56$ ($\mu$m),
$\epsilon_F\approx 84.3$ ($\hbar\omega_r$), $k_F\approx 9.2/a_r $,
and $n(\vec r=0)=7.45\times 10^{13}$/cm$^3$.
}
\label{fig2}
\end{figure}

\begin{figure}
\caption{The same as for Fig. 2, but now for $^{82,84,86}$Rb with
$a_r\sim 0.39$ ($\mu$m)
and $n(\vec r=0)\sim 2.1\times 10^{14}$/cm$^3$. All
three isotopes give similar results since hyperfine structures
are not included here. The half lifetimes of the three isotopes are
1.25 minute, 32.9, and 18.8 days respectively.
}
\label{fig3}
\end{figure}


\begin{references}
\bibitem{noteco2}For the CO$_2$ laser at a wavelength of $\lambda_L=10.6$
($\mu$m), the strength of the quasi-static electric field is given by
${\cal E}\sim 27 \sqrt{I}$ ${\rm(V/cm)}$, with laser intensity
$I$ in units of (watts/{cm$^2$}). Therefore a field of $10^6$ (V/cm)
requires $I\sim$ $10^9$ (watts/cm$^2$).
\end{references}
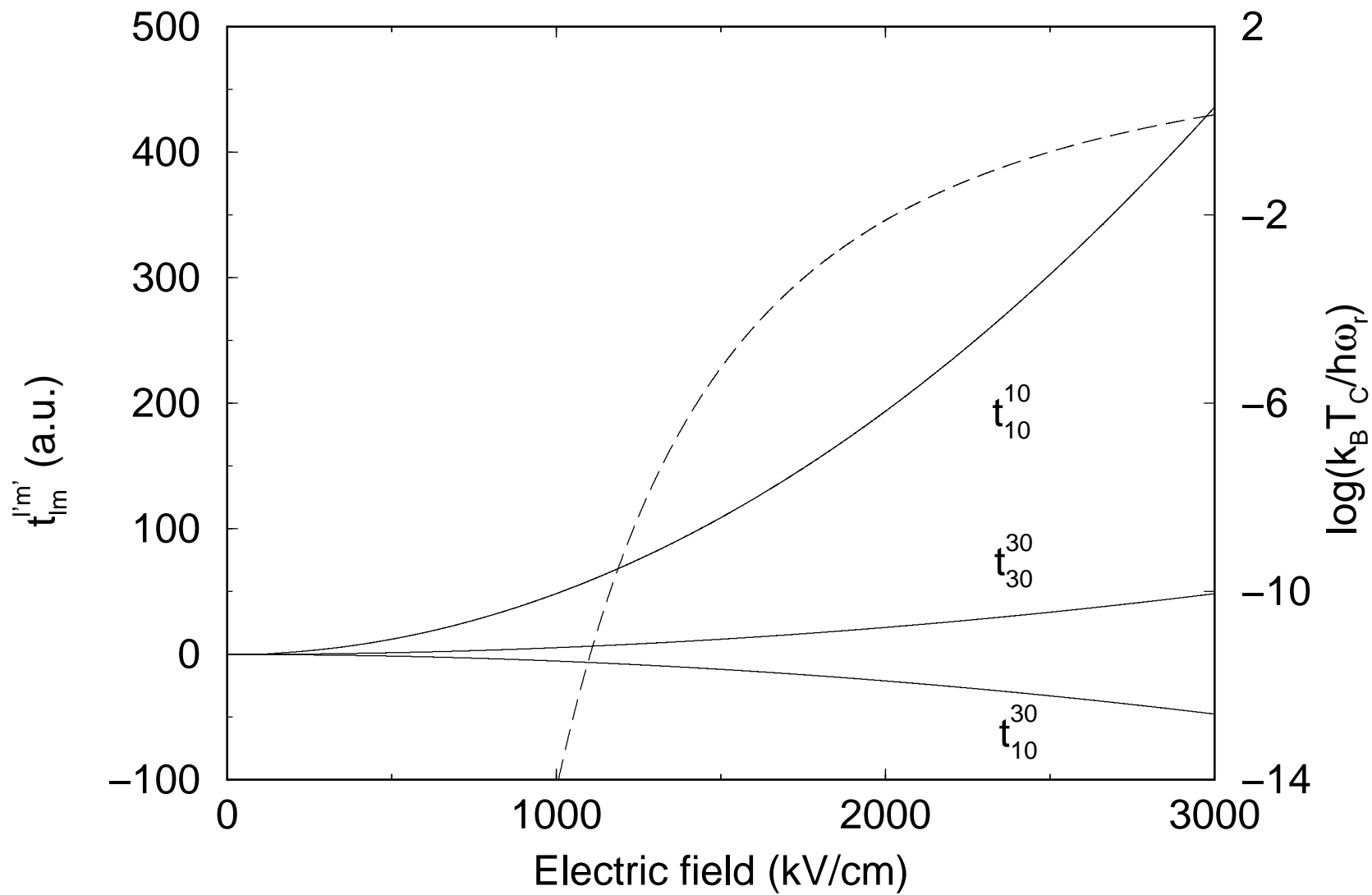
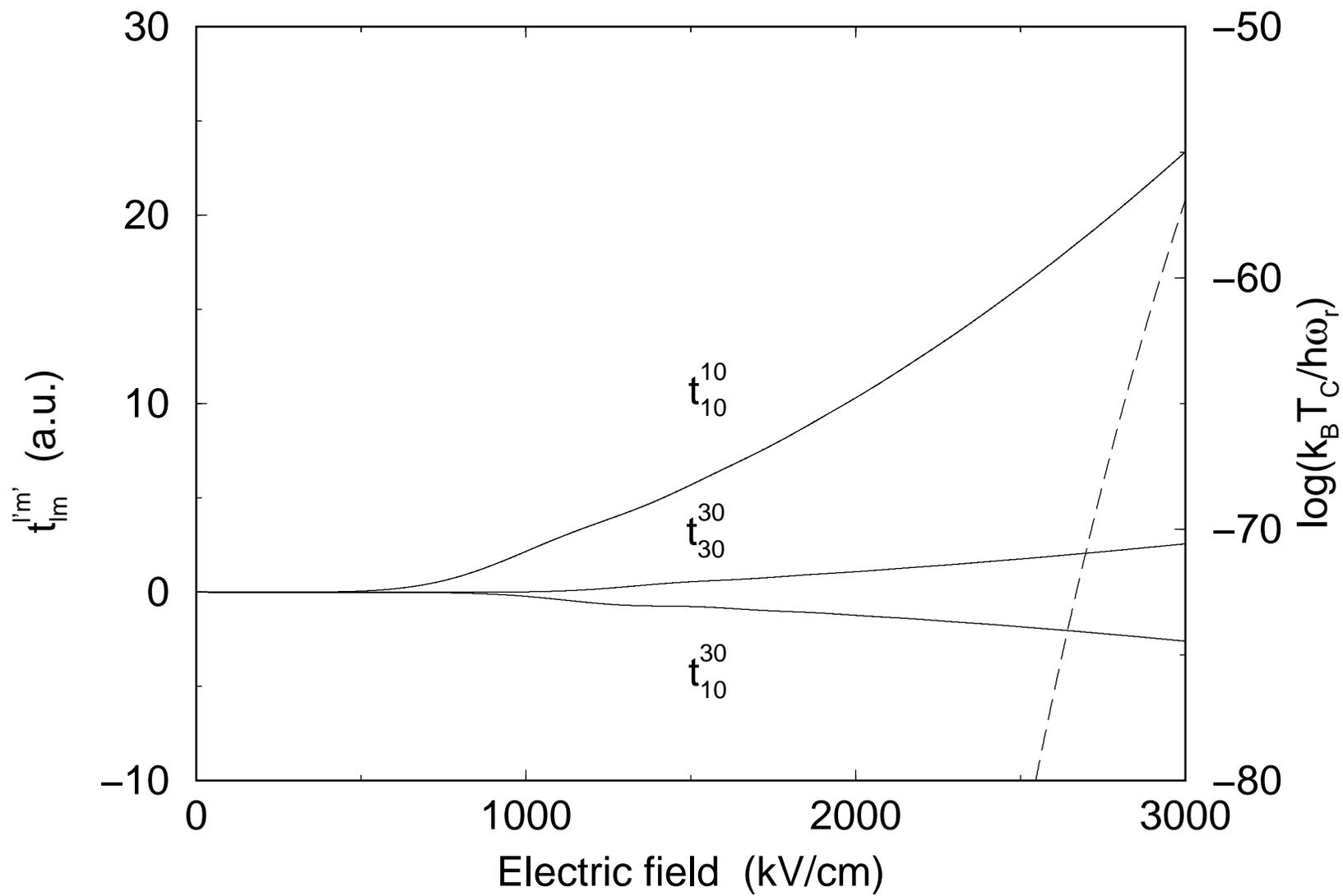
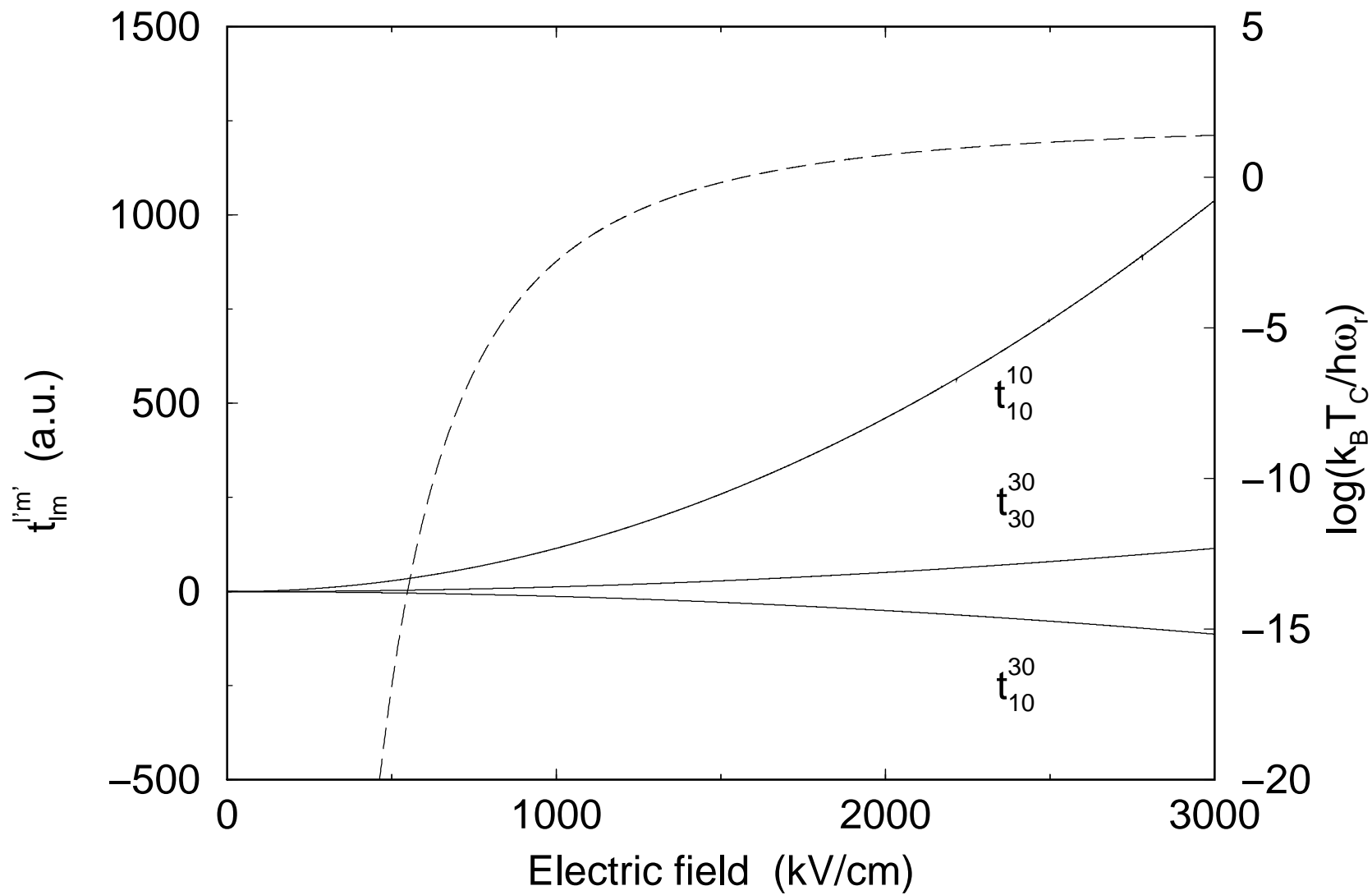
\end{document}